\documentclass[12pt]{article}
\newlength{\vshift}
\input epsf.tex
\newlength{\hshift}

\usepackage{amssymb}
\usepackage{amsmath}
\usepackage{amsthm}
\usepackage{amsopn}
\def\uno{\mbox{1 \kern-.59em {\rm l}}}
\def\la{\lambda}
\def\La{\Lambda}
\def\ka{\kappa}
\def\de{\delta}
\def\p{\partial}

\def\th{\theta}
\def\nn{\nonumber}
\def\be{\begin{equation}}
\def\ee{\end{equation}}
\def\bea{\begin{eqnarray}}
\def\eea{\end{eqnarray}}

\begin{document}

 \vspace*{3cm}

 \begin{center}

 {\bf{\Large Singlet particles as cold dark matter in noncommutative space-time}}

\vskip 4em

 { {\bf M. M. Ettefaghi} \footnote{ ettefaghi@qom.ac.ir  }}
 \vskip 1em
 Department of Physics, The University of Qom, Qom 371614-611, Iran.

 \end{center}

 \vspace*{1.9cm}

\begin{abstract}
 We extend the noncommutative (NC) standard model to incorporate singlet particles as cold dark matter. In the NC space-time, the singlet particles can be coupled to the $U(1)$ gauge field in the adjoint representation. We Study the relic density of the singlet particles due to the NC induced interaction. Demanding either the singlet fermion or the singlet scalar to serve as cold dark matter and the NC induced interactions to be relevant to the dark matter production, we obtain the corresponding relations between the NC scale and the dark matter masses, which are consistent with some existing bounds.
\end{abstract}

\newpage
\section{Introduction}
 Although no new physics beyond the standard model (SM) appeared so far in collider experiments, the dark matter (DM) problem as well as neutrino oscillations, baryon asymmetry, and so on calls for new physics beyond the SM. In fact, looking for a new model beyond the SM to provide a candidate of DM seems to be natural. Supersymmetry models with R parity \cite{susy,PDM}, the extra dimensional models with conserved KK parity \cite{extra}, the T-parity conserved little Higgs model \cite{little} and so on have the same capability. Especially, it is also possible to add singlet fermions or singlet scalars to the SM to serve as cold DM. Namely, \begin{itemize}

            \item
            The most minimal model, which include a singlet fermion to serve for dark matter, is discussed in \cite{minimalmodel}. In this model the leading interaction between dark matter and standard model particle is given by the dimension five term $\frac{1}{\La}H^\dag H\bar{\psi}\psi$, where $H$, $\psi$ and $\La$ are, respectively, the SM Higgs doublet, the DM fermion and a new physics scale. Moreover, it is possible to explain the dark matter production in the early universe by a renormalizable theory if the Higgs sector of the SM is extended by a new singlet Higgs \cite{sigletfermion}.

            \item
            As another scenario to have a singlet DM, the SM is extended by adding some singlet scalars \cite{scalar}. In this group of models the DM production is explained by a coupling between the SM Higgs and the DM. Here it is necessary to define a $Z_2$ symmetry under which the SM particles are even and the DM particles are odd, to guarantee the stability of the DM.
          \end{itemize}

On the other hand, considering quantum mechanics together with classical general relativity leads to the existence of a minimal length in the nature \cite{minimallength}. Therefore, the noncommutative (NC) field theories as simple examples of models with minimal length are very interesting. Moreover, the NC field theories also have been motivated from the string theory, for a review see \cite{douglas}. Many authors have recently studied the NC field theories extensively \cite{nNCSM,ncg,sm,anomaly,renormal,neutral boson,pho-neu,hez,eh,charge,phenomenology,nucleosynthesis}. The NC field theories are constructed on space-time coordinates which are assumed as operators and do not obey commutative algebra. In the case of canonical version of the NC space-time, the coordinates obey the following algebra:
\be
\th^{\mu\nu}=-i[\hat{x}^\mu,\hat{x}^\nu],
\ee
where a hat indicates a NC coordinate and $\th^{\mu\nu}$ is a real, constant, and antisymmetric matrix. To construct the NC field theory, according to the Weyl-Moyal correspondence, an ordinary function can be used instead of the corresponding NC one by replacing the ordinary product with the star product as follows:
\be\label{starproduct}
 f\star
g(x,\theta)=f(x,\theta)\exp(\frac{i}{2}\overleftarrow{\partial}_\mu
\theta^{\mu\nu}\overrightarrow {\partial}_\nu)g(x,\theta).
\ee
Using this correspondence, however, there are two approaches to
construct the gauge theories in the NC space. In the
first one, the gauge group is restricted to $U(n)$ and the symmetry
group of the standard model is achieved by the reduction of
$U(3)\times U(2)\times U(1)$ to $SU(3)\times SU(2)\times U(1)$ by an
appropriate symmetry breaking \cite{nNCSM}. In the second approach,
the NC gauge theory can be constructed for $SU(n)$ gauge
group via Seiberg-Witten map \cite{ncg}. Therefore, in this manner, we can write the NC SM by replacing the usual products in the Lagrangian by the star-products and by writing the NC fields with respect to the usual fields using the Seiberg-Witten maps \cite{sm}. While it is necessary to introduce two new gauge fields and two new Higgs fields in the former, the number of the degrees of freedom in the latter is as many as the usual SM. Moreover, we have to emphasize that the latter version of NC SM is anomaly free \cite{anomaly}, and the renormalization properties of the gauge theory based on Seiberg-Witten maps have already been considered \cite{renormal}.
However, in both versions of the NC SM, if a gauge singlet field is added, it achieves an NC induced coupling to $U(1)$ gauge field similar to the neutral bosons-photon interaction in NC QED \cite{neutral boson}, the photon-neutrino interaction in NC QED constructed using Seiberg-Witten map \cite{pho-neu}, and the $U(1)$ gauge boson-right handed neutrino interaction in the NC SM based on $SU(3)\times SU(2)\times U(1)$ gauge groups \cite{hez}. In the case of neutrinos, this problem causes not to be possible to write Yukawa terms for the neutrinos since they violate the gauge symmetry \cite{eh}. In addition, considering NC corrections in the SM using Seiberg-Witten maps leads to other new physics; this has been studied extensively in the literature such as \cite{phenomenology}. Especially, a constraint on the NC scale has been obtained by studying the early universe in the NC background \cite{nucleosynthesis}.

In this paper, we extend the NC SM by adding a singlet fermion and/or a singlet scalar in a gauge invariant manner. Then, we study the relic abundance of them which is due to the NC induced interactions. This paper is organized as follows.  In the next section, we review the singlet DM models. In the third section, at first we summarize the NC SM; then we discuss on the extension of the NC SM to include the singlet particles. In the forth section, we study the relic abundance of the singlet particles due to the NC induced interactions. Also we discuss on the consistent conditions under which either singlet particles plays the cold DM role. Finally, we discuss on our conclusions in the last section.
\section{Review of singlet dark matter models}
The SM does not predict any appropriate candidate to serve as DM. So it has been tried to introduce a proper candidate for the DM in the many extensions of the SM. Especially, in the simplest extension, some singlet particles are added to the SM content to serve as cold DM. Here, we briefly introduce them during the two following items:
\begin{itemize}
            \item
            It is possible to consider a singlet Dirac fermion beyond the SM which can serve as cold dark matter\cite{minimalmodel}. The lepton number and baryon number of this new fermion is taken to be zero. Hence, there is not any mixing between the new fermion and the standard model fermions. However, it is possible to have an interaction between this particle and SM particles through the following dimension five term:
            \be
            {\cal{L}}_{int}=-\frac 1 \La H^{\dag}H\bar{\psi}\psi
            \ee
            where $\La$, $H$ and $\psi$ are a new scale, the SM Higgs doublet and the DM candidate, respectively. However, we can have an interaction between the SM particles and new singlet fermion in a renormalizable manner by adding a singlet scalar, $S$, too \cite{sigletfermion}. The Lagrangian of this theory is written as follows:
            \be
            {\cal L}={\cal L}_{SM}+{\cal L}_{hid}+{\cal L}_{int},\ee
            where the hidden sector lagrangian is given by
            \be
            {\cal L}_{hid}={\cal L}_{S}+{\cal L}_{\psi}-g_S\bar{\psi}\psi S,\ee
            with
            \be
            {\cal L}_S=\frac 1 2 \p_\mu S\p^\mu S -\frac{m_0^2}{2}S^2-\frac{\la_3}{3!}S^3-\frac{\la_4}{4!}S^4, \nn \ee
            \be
            {\cal L}_{\psi}=\bar{\psi}(i\p\!\!\!/-{m_X}_0)\psi.
            \ee
            The interaction Lagrangian between the hidden sector and the SM fields is given by
            \be
            {\cal L}_{int}=-\la_1H^\dag H S-\la_2H^\dag H S^2,
            \ee
            where $H$ is the SM Higgs doublet whose potential is $-\mu^2H^\dag H+\bar{\la_0}(H^\dag H)^2$. The vacuum expectation values of Higgs doublet are
            \be
            \langle H\rangle=\frac 1 {\sqrt{2}}\left(
                                                 \begin{array}{c}
                                                   0 \\
                                                   v_0 \\
                                                 \end{array}
                                               \right),
            \ee
            and for the singlet fermion is $\langle S\rangle=x_0$. The Higgs scalar fields which appear above are not in the mass eigenstate. Therefor, writing the Higgs field in the mass eigenstate, we find that the interactions of Singlet fermion and SM particles are suppressed by the mass of singlet scalar and/or Higgs mixing parameter. Hence, $\psi$ is naturally weakly interacting massive particle (WIMP).
            After symmetry breaking, the mass of singlet fermion is $m_X={m_X}_0+g_Sx_0$. Clearly, it is a independent parameter of the model since ${m_X}_0$ is a free parameter. Then, we can find a compatible region in the parameter space where $\psi$ plays the cold DM role \cite{sigletfermion}.
            \item
            The other possibility to have singlet DM is by adding some singlet scalars to the SM \cite{scalar, burgess}. The Lagrangian of this model is
            \be
            {\cal L}={\cal L}_{SM}+\frac 1 2\p_\mu S\p^\mu S-\frac{m_0^2}{2}S^2-\frac{\la_s}{4}S^4-\la S^2H^\dag H,
            \ee
            where there exist three parameters, $m_0$, $\la_s$ and $\la$, in addition to the SM. The coupling to all SM fields are controlled by the single parameter $\la$. Meanwhile, the scalar potential of the model in the unitary gauge $H=\frac 1 {\sqrt{2}}\left(
                                                 \begin{array}{c}
                                                   0 \\
                                                   h \\
                                                 \end{array}
                                               \right)
            $ is given by
             \be
            V=\frac{m_0^2}{2}S^2+\frac{\la_s}{4}S^4+\frac{\la}{2}S^2h^2+\frac{\la_h}{4}(h^2-v_0^2)^2.
            \ee
            To bound the potential from below, we consider
            \bea
            &&\la_s,\,\la_h\geq0\,\,\,\,\mbox{and}\nn\\
            &&\la_s\la_h\geq \, \la^2\,\,\,\mbox{for negative $\la$}.
            \eea
            Moreover, to have acceptable particle masses and to guarantee the stability of $S$, we demand the vacuum expectation value of $h$, $\langle h\rangle$, and of $S$, $\langle S\rangle$, to be, respectively, nonzero and  zero. This conditions occur if $v_0^2>0$ and consequently $\langle h\rangle=v_0^2$ \cite{burgess}. After replacing $h$ by $h+v_0$, the $S$-dependent part of the scalar potential is given by
            \be
            V=\frac{1}{2}(m_0^2+\la v_0^2)S^2+\frac{\la_s}{4}S^4+\frac{\la}{2}S^2h^2+\la v_0S^2h,
            \ee
            in which $h$ measures the deviations of the $H$ from the equilibrium ground state configuration, $v_0$. Therefore, the $S$ mass is a free parameter too and it can be constrained such that it plays the cold DM role.
          \end{itemize}
\section{Extension of noncommutative standard model by singlet sectors}
 There are two approaches to construct the SM model in the NC space-time.

 In one of these, the usual SM gauge symmetry is obeyed and the number of gauge fields, couplings, and particles are the same as the ordinary one. However, there exist some new
interactions induced by the NC space-time in comparison to the usual SM \cite{sm}. The phenomenological aspects of this new interaction have been considered extensively in the literature \cite{phenomenology}.
We denote the fermion content of the theory as
 \be
  \hat{L}=\left(
\begin{array}[]{c}
\hat{\Psi}_{L_{\nu_l}}\\ \hat{\Psi}_{L_l}
\end{array}
\right)\,\,\,\, , \,\,\,\,\left( \begin{array}[]{c}
\hat{\Psi}_{L_u}\\ \hat{\Psi}_{L_d}
\end{array} \right)  \ee
and \bea
\hat{R}\!\!\!\!&=&\!\!\!\!\hat{\Psi}_{R_l}\,\,\,\, ,
\,\,\,\,\hat{\Psi}_{R_u}\,\,\,\, , \,\,\,\
\hat{\Psi}_{R_d},
\eea
where for three generations, $l$ stands for $e$, $\mu$ and $\tau$ meanwhile subscript $u$
refers to up-type quarks and subscript $d$ to down-type quarks. The
fields
 with hat are the non-commutative fields which can be written as a
 function of ordinary fields using appropriate SW-maps
 \be
 \hat{\Psi}(V_\mu)=\Psi^0+\frac{1}{2}\th^{\mu\nu}V_\nu\p_\mu\Psi+
 \frac{i}{8}\th^{\mu\nu}[V_\mu,V_\nu]\Psi+{\cal{O}}(\th^2),\ee
where $$V_\mu=gT^aW^a_\mu+g'\frac{Y}{2}B_\mu+g_sT^a_sG_s.$$
Here $T^a_s$, $T^a$ and $\frac{Y}{2}$ are the generators of
the gauge groups $SU_C(3)$, $SU_L(2)$ and $U_Y(1)$, respectively, and
have to consider in appropriate representation according to the
representation of fermions, introduced in table (1). The Lagrangian of the fermionic sector is
\be S_{fermion}=\int
d^4x(\overline{\hat{L}}\star i\hat{{\cal
D}}\!\!\!\!/\,\hat{L}+\overline{\hat{R}}\star
i\hat{{\cal
D}}\!\!\!\!/\,\hat{R}), \ee
where
\be \label{cod} \hat{{\cal
D}}_\mu=\p_\mu-igT^a\hat{W}^a_\mu-ig'\frac{Y}{2}\hat{B}_\mu-ig_sT^a_s\hat{G_s}.
\ee
 The Seiberg-Witten
maps of gauge bosons can been written as \be\label{v}
\hat{V}_\mu=V_\mu+\frac{1}{4}\th^{\ka\la}\{\p_\ka
V_\mu,V_\la\}+\frac{1}{4}\th^{\ka\la}\{F_{\ka\mu},V_\la\}+{\cal{O}}(\th^2).\ee

In the other approach, the gauge group is $U_\star(3)\times U_\star(2)\times U_\star(1)$ which is reduced to $SU(3)\times SU(2)\times U(1)$ by reducing the two extra $U(1)$ factors through the appropriate Higgs mechanism and Higgs particles (Higgsac) \cite{nNCSM}. In this approach the number of possible particles in each family( which are six: left-handed lepton, right-handed charged lepton, left-handed quark, right-handed up quark, right-handed down quark, and and Higgs boson) as well as their hypercharge are naturally fixed. But it is possible to add every singlet particle to the mentioned set.

According to the Weyl-Moyal correspondence, (\ref{starproduct}), the usual products in $gA\phi$, where $g$, $A$ and $\phi$ are coupling, gauge field and matter field, respectively, are replaced by star products and it leads to an ambiguity in the ordering of fields: $gA\star\phi$, $g\phi\star A$ and $g(A\star\phi-\phi\star A)$. In the action, however, it has been shown that the two first coupling are the charge conjugation of each other but the third one is the charge conjugation of itself \cite{charge}. Therefore the gauge singlet particle added to the SM content can have the third coupling.

It must be noticed that in the former approach if the singlet fermion is right-handed neutrinos, Yukawa coupling among the left-handed and right-handed neutrinos and the SM Higgs field is not possible because of the violation of the gauge symmetry \cite{eh}. Otherwise, if a singlet fermion, $\psi$, is added to the SM, the following term in the NC space-time is possible
\be\label{psi}
\de{\cal L}=\bar{\hat{\psi}}\star
i\hat{{\cal D}}\!\!\!\!/\,\hat{\psi}-m_X\bar{\hat{\psi}}\star\hat{\psi},
\ee
where
\be \label{adjoint}\hat{{\cal
D}}_\mu=\p_\mu-ig^\prime[\hat{B}_\mu,\hat{\psi}]_\star, \ee
and
\be\label{SW2}\hat{\psi}=\psi+g^\prime\th^{\mu\nu}B_\mu\p_\nu\psi.\ee
 Moreover, if a singlet scalar, $S$, is considered in addition to the SM content too, in the NC space-time, the following term can be added to the NC SM
\be\label{s}
\de{\cal L}=\hat{{\cal D}}{\hat{S}}\star
\hat{{\cal D}}\,\hat{S}-m_X^2\hat{S}\star\hat{S}\,,
\ee
in which $\hat{S}$ and $\hat{{\cal D}}$ are similar to (\ref{adjoint}) and (\ref{SW2}), respectively. The gauge transformation of the singlet both fermion and scalar is as follows
\be
\de\hat{\phi}=i\hat{\Lambda}^\prime\star\hat{\phi}-i\hat{\phi}\star\hat{\Lambda}^\prime \,,
\ee
where $\phi$ stands for both $\psi$ and $S$ and $\hat{\Lambda}^\prime$ is the gauge parameter of $U_Y(1)$. So it is clear that the Yukawa coupling between the singlet fermion and the singlet scalar,  $\overline{\hat{\psi}}\star\hat{S}\star\hat{\psi}$, is gauge invariant. In addition, the interaction terms between $S$ and the standard model Higgs doublet, $H$, such as
 $H^\dag HS$ and $H^\dag HS^2$ do not violate the gauge symmetry, if $H$ transforms as follows
 \be
 \de\hat{H}=i\hat{\Lambda}\star\hat{H}-i\hat{H}\star\hat{\Lambda}^\prime
 \ee
 in which $\Lambda$ is the gauge parameter of $SU(2)$.
\begin{table}
\begin{center}
\begin{tabular}
 {|c|c|c|c|}\hline
    & $SU(3)_c$ & $SU(2)_L$ & $U(1)_Y$ \\ \hline
  $e_R$ & 1 & 1 & -2 \\ \hline
  $L_l=\begin{pmatrix}
    \nu_L \\
    e_L \
  \end{pmatrix}$ & 1 & 2 & -1 \\ \hline
  $u_R$ & 3 & 1 & $\frac{4}{3}$ \\ \hline
  $d_R$ & 3 & 1 & $\frac{-2}{3}$ \\ \hline
  $L_q=\begin{pmatrix}
    u_L \\
    d_L \
  \end{pmatrix}$ & 3 & 2 & $\frac{1}{3}$ \\ \hline
  $\Phi=\begin{pmatrix}
    \phi^+ \\
    \phi^0 \
  \end{pmatrix}$ & 1 & 2 & 1 \\ \hline$\nu_R$ & 1 & 1 & 0 \\ \hline
\end{tabular}
\end{center}
\caption{Matter and Higgs field content of the extended standard
model and their representation}
\end{table}

In the latter approach, the singlet fields and $U_\star(1)$ gauge field transform as follows
\bea
&& B_\mu\rightarrow v\star B_\mu\star v^{-1}+\frac i {g^\prime}v\star\p_\mu v,\nn\\
&& \phi\rightarrow v\star\phi \star v^{-1},
\eea
in which $v$ is the $U_\star (1)$ element. Hence, the interactions (\ref{psi}) and (\ref{s}) as well as $\bar{\psi}S\psi$ can be written similarly in the gauge invariant manner. However, since $H\rightarrow V\star H$ where $V$ is the $U_\star(2)$ element, the interactions between $H$ and $S$ such as $H^\dag HS$ and $H^\dag HS^2$ are not gauge invariant. Therefore, in the NC SM based on the $U(3)\times U(2)\times U(1)$ gauge group, the interaction between the SM Higgs and the singlet scalar is not possible. However, if there exist any singlet scalar particle, we may explain its production in the early universe by the NC induced interaction. We discuss it in the following section.
\section{Singlet dark matter in the noncommutative space}
When the interaction rate of a particle species in the early universe drops below the expansion rate of the universe, it falls out of thermodynamics equilibrium and its number density in the comoving volume remains constant. The relic density of a generic weakly interacting massive particle (WIMP) is estimated as follows \cite{PDM}
\be
\Omega_Xh^2\approx\frac{3\times10^{-27}cm^3s^{-1}}{\langle\sigma_{ann} v\rangle},
\ee
where the normalized abundance $\Omega_X$ is defined as $\Omega_X=\rho_X/\rho_c$, where $\rho_X$ and $\rho_c$ are the WIMP density and critical density respectively, and the scaled Hubble parameter $h$ is defined as $H_0\equiv100h \,\mbox{km}\,\mbox{s}^{-1}\mbox{Mpc}^{-1}$. $\langle\sigma_{ann} v\rangle$ is the thermal average of the cross section times velocity. According to recent cosmological observations, we know that the normalized abundance of the DM is about $0.22$ \cite{density}. Therefore, if $X$ is a DM candidate, the thermal average of its cross section must satisfy the following constraint
\be\label{c}
\langle\sigma_{ann} v\rangle\sim1.4\times10^{-26} \mbox{cm}^3\,\mbox{s}^{-1}\simeq1.2\times10^{-9}\,GeV^{-2}.
\ee
 For non-relativistic gases, excluding the regions that $\sigma v_{ann}$ varies rapidly with $\epsilon=\frac{s-4m^2_x}{4m^2_x}$, as near the thresholds, we can obtain an expansion for $\langle\sigma_{ann} v\rangle$ in power of $x^{-1}\equiv T/m_x$, where $T$ is the freeze-out temperature \cite{avar}. This expansion, up to the second order of $x^{-1}$, is given by:
\be\label{average}
\langle\sigma_{ann} v\rangle\simeq a^{(0)}+\frac 3 2 a^{(1)}x^{-1}+\frac{15}{8} a^{(2)}x^{-2},
\ee
in which $a^{(k)}$ indicates the $k$th derivative of $\sigma_{ann} v$ with respect to $\epsilon$ evaluated at $\epsilon=0$. The freeze-out temperature of weakly interacting massive particle is about $T\simeq 25m_x$.

 As was shown, in the NC space, a singlet particle as well as the other particles of the SM can be involved in the gauge interactions.
 Therefore, beside the other considered interactions, these new interactions may be also relevant in the relic abundance of the singlet DM. Besides that, the considered interactions to explain the singlet particles production in the usual space-time are not consistent with gauge symmetry in the NC SM based on $U(3)\times U(2)\times U(1)$ gauge group. The most dominated contribution of the NC interactions is related to the photon exchange annihilation processes. At the leading order, we have to consider the annihilation of the singlet DM to the charged fermion and antifermion and to $W^+W^-$. If we take $\th^{i0}=0$, these annihilation cross section are zero. Otherwise, we have non zero contribution if $\th^{i0}\neq0$. We define the vector $\overrightarrow{\th}$ as follows
 \be
 \overrightarrow{\th}=(\th^{01},\th^{02},\th^{03}).
 \ee
 When we expand the NC action with respect to $\th$, the order of magnitude of the measurable quantities calculated in either version of the NC SM are not different from the other version. Therefore, we use Seiberg-Witten maps method to write the related action for our purpose without missing generality.  The related action to coupling of the singlet fermion to hyper-photon in the NC space-time can be written using (\ref{psi}), (\ref{v}), (\ref{SW2}), and (\ref{starproduct}) as follows
 \be
 S_{\psi\gamma}=-\frac{g^\prime}{2}\th^{\nu\rho}\int{d^4x\bar{\psi}
 (i\gamma^\mu(B_{\nu\rho}\p_\mu+B_{\mu\nu}\p_\rho+B_{\rho\mu}\p_\nu)-m_XB_{\nu\rho})}.
 \ee
 Similarly, in the case of singlet scalar, using (\ref{s}), ({\ref{v}}), (\ref{SW2}), and (\ref{starproduct}), it is written as follows
 \be\label{scalaraction}
 S_{S\gamma}=-g^\prime\th^{\mu\nu}\int{d^4x(\p^2SB_\nu\p_\mu S-\p_\nu S\p_\mu B_\rho \p_\rho S+m_X^2SB_\nu\p_\mu S}).
 \ee
 $B_\mu$ in the mass eigenstate is given by
 \be
 B_\mu=\cos\th_WA_\mu-\sin\th_W{Z_0}_\mu,
 \ee
 and $g^\prime=e\cos\th_W$.

 Now we are ready to calculate approximately the thermal average annihilation cross section of the singlet particles.
\begin{itemize}
\item
 In the case of singlet fermion, $a^{(0)}$ is zero but $a^{(1)}$ is nonzero. Therefore, the NC contribution of the thermal average annihilation cross section of the singlet fermion is estimated as follows
\bea\label{fa}
{\langle\sigma_{ann} v\rangle}_{\mbox{fermion}}\simeq\pi^4\th^2\Big(&&\!\!\!\!\!\!\sum_f\frac{640m_X^2+256m_f^2}{15015200}\sqrt{1-\frac{m_f^2}{m_X^2}}
\nn \\&&+\frac{14(m_X^2-m_W^2)}{1876900}\sqrt{1-\frac{m_W^2}{m_X^2}}\Big),
\eea
where the first term is due to the annihilation to the pair charged fermions whose masses are below the singlet fermion mass and the second term exists if the singlet fermion mass is larger than $W^\pm$ mass. However, the contribution of the first term in (\ref{fa}) is more than the second one. Meanwhile, the contributions of the allowed various pair charged fermions are in the same order. Therefore, we have
\be
{\langle\sigma_{ann} v\rangle}_{\mbox{fermion}}\sim 3.9\times10^{-3}N_f\frac{m_X^2}{\La_{NC}^4},
\ee
where $\La_{NC}\equiv\frac 1 {\sqrt{\th}}$ and $N_f$ denotes the number of allowed pair charged fermion. If the singlet fermion describes the DM, using (\ref{c}) we find that the NC induced interactions are relevant to the DM production in early universe provided that the NC scale and the DM mass satisfy the following relation
\be
\La^4_{NC}\sim3.2\times10^6N_fm^2_X.
\ee
For instance, if $m_X$ is of the order of $100\,GeV$ and the NC scale is in the order of $600\,GeV$, which are consistent with the existing bounds, the recent model can be employed.
\item
Using (\ref{average}) to obtain approximately the thermal average annihilation cross section of the singlet scalars, we find that both $a^{(0)}$ and $a^{(1)}$, which are due to (\ref{scalaraction}), are zero. However, the next order contribution, $a^{(2)}$, in the leading order with respect to the NC parameter, is nonzero and its order of magnitude is more than the order of magnitude of $a^{0}$ and $a^{(1)}$ up to the second order of NC parameter. Hence, the thermal average of the annihilation cross section of the singlet scalars is obtained as follows
\bea\label{sa}
{\langle\sigma_{ann} v\rangle}_{\mbox{scalar}}\simeq\pi^4\th^2\Big(&&\!\!\!\!\!\!\sum_f\frac{m_X^2+m_f^2}{5005}\sqrt{1-\frac{m_f^2}{m_X^2}}
\nn \\&&+\frac{(m_X^2-m_W^2)}{1042722}\sqrt{1-\frac{m_W^2}{m_X^2}}\Big).
\eea
Here the second term, which is due to the annihilation to $W^\pm$, is suppressed by a factor of the order of $10^{-3}$. Therefore, we can estimate the thermal averaged annihilation cross section of singlet scalars as follows
\be
{\langle\sigma_{ann} v\rangle}_{\mbox{scalar}}\sim 1.9\times10^{-2}N_f\frac{m_X^2}{\La_{NC}^4}.
\ee
Consequently, if the singlet scalar serve as cold DM, using (\ref{c}) we find that if the NC scale and the DM mass satisfy the following relation
\be
\La_{NC}^4\sim1.6\times10^7N_fm_X^2,
\ee
the NC induced interactions are relevant to the DM production. For instance, if $m_X$ is in the order of $100GeV$ and the NC scale is in the order of $1TeV$, which are consistent with the existing bounds, this model also can be employed.
\end{itemize}
\section{conclusion}
In this paper we have extended both versions of the NC SM, the NC SM based on the $SU(3)\times SU(2)\times U(1)$ gauge group and the NC SM based on the $U(3)\times U(2)\times U(1)$ gauge group, by gauge singlet sectors. In the usual space-time, this additional sectors can interact with the SM particles through coupling to the SM Higgs. It is possible to have some allowed regions in the parameter space where the singlet particles serve to describe the cold DM \cite{minimalmodel,sigletfermion,scalar,burgess}. In the NC space-time, the gauge singlet particles can have interaction directly with the $U(1)$ gauge field in the adjoint representation; therefore, the singlet particles can be involved in the gauge transformation. As a result, to conserve the gauge symmetry, the usual interactions between the singlet sectors and the SM Higgs can be only transcribed to the NC SM based on the $SU(3)\times SU(2)\times U(1)$ gauge groups. Then, we have studied the relic density of the singlet particles which is due to the NC induced interactions. Demanding one of the gauge singlet particles to serve as cold DM and considering the consistency with the existing bounds on the NC scale and the DM mass, we have shown the NC induced interactions can be relevant to the DM production.

{\bf Acknowledgment}
The author would like to thank H. Razmi for useful discussion and reading the manuscript.


\begin{thebibliography}{99}
\bibitem{susy}
G. Jungman, M. Kamionkowski and K. Griets, Phys. Rep. {\bf 267}, 195 (1996).
\bibitem{PDM}
G. Bertone, D. Hooper, J. Silk, Phys. Rept. {\bf 405}, 279 (2005).
\bibitem{extra}
H.-C. Cheng, J.L. Feng and K.T. Matchev, Phys. Rev. Lett. {\bf 89}, 211301 (2003); G. Servant and T. Tait, Nucl. Phys. B{\bf 650}, 391 (2003).
\bibitem{little}
H.-C. Cheng and I. Low, JHEP {\bf 0309}, 051 (2003); JEHP {\bf 0408}, 061 (2004).
\bibitem{minimalmodel}
Y.G. Kim, and K.Y. Lee, Phys. Rev. D{\bf 75}, 115012 (2007).
\bibitem{sigletfermion}
Y.G. Kim, K.Y. Lee, and S. Shin, JHEP {\bf 0805}, 100 (2008).
\bibitem{scalar}
J. McDonald, Phys. Rev. D {\bf 50}, 3637 (1994) [arXiv:hep-ph/0702143];
 J. McDonald, Phys. Rev. Lett. {\bf 88}, 091304 (2002)
[arXiv:hep-ph/0106249];  M.
C. Bento, O. Bertolami and R. Rosenfeld, Phys. Lett. B{\bf 518}, 276
(2001) [arXiv:hep-ph/0103340]; H. Davoudiasl, R. Kitano, T. Li and
H. Murayama, Phys. Lett. B{\bf 609}, 117 (2005) [arXiv:hep-ph/0405097].
\bibitem{burgess}
C. P. Burgess, M. Pospelov and T. ter
Veldhuis, Nucl. Phys. B{\bf 619}, 709 (2001) [arXiv:hep-ph/0011335];
\bibitem{minimallength}
X. Calmet, M. Graesser, and S.D.H. Hsu, Phys. Rev. Lett. {\bf 93}, 211101 (2004).
\bibitem{douglas}M.R. Douglas and N.A. Nekrasov, Rev. Mod. Phys. {\bf 73}, 977{2002}.
\bibitem{nNCSM} M. Chaichian, P. Presnajder, M.M. Sheikh-Jabbari and A.
Tureanu, Eur. Phys. J. C{\bf 29}, 413(2003).
\bibitem{ncg}
J.~Madore, S.~Schraml, P.~Schupp and J.~Wess, Eur.\ Phys.\ J.\ C{\bf 16}, 161 (2000) [hep-th/0001203]; B.~Jur\v co, S.~Schraml,
P.~Schupp and J.~Wess, Eur.\ Phys.\ J. C{\bf 17}, 521 (2000)
[hep-th/0006246]; B.~Jur\v co, L.~M\"oller, S.~Schraml, P.~Schupp
and J.~Wess, Eur.\ Phys.\ J. C{\bf 21}, 383 (2001) [hep-th/0104153];
Lutz~M\"{o}ller, JHEP {\bf 10}, 063 (2004).
\bibitem{sm}
X.~Calmet, B.~Jur\v co, P.~Schupp, J.~Wess and M.~Wohlgenannt, Eur.\
Phys.\ J. C{\bf 23}, 363 (2002) [hep-ph/0111115]; B. Meli\'{c}, K.
Passek-Kumeri\v{c}ki, J. Trampeti\'{c}, P. Schupp and M.
Wohlgenannt, Eur.\ Phys.\ J. C{\bf 42}, 483(2005).
\bibitem{anomaly}
C.~P.~Martin, Nucl.\ Phys.\  B{\bf 652} (2003) 72 [arXiv:hep-th/0211164];
F.~Brandt, C.~P.~Martin and F.~R.~Ruiz, JHEP {\bf 0307}, 068 (2003)
[arXiv:hep-th/0307292].
\bibitem{renormal}
J.~M.~Grimstrup and R.~Wulkenhaar, Eur.\ Phys.\ J.\  C{\bf 26}, 139
(2002)[arXiv:hep-th/0205153]; A.~Bichl, J.~Grimstrup, H.~Grosse, L.~Popp,
M.~Schweda and R.~Wulkenhaar, JHEP {\bf 0106} (2001) 013
[arXiv:hep-th/0104097];  M.~Buric, V.~Radovanovic and J.~Trampetic, JHEP {\bf 0703} (2007) 030 [arXiv:hep-th/0609073]; D.~Latas,
V.~Radovanovic and J.~Trampetic, Phys.\ Rev.\  D{\bf 76} (2007) 085006
[arXiv:hep-th/0703018]; C.~P.~Martin, D.~Sanchez-Ruiz and
C.~Tamarit, JHEP {\bf 0702}, 065 (2007); C.~P.~Martin and C.~Tamarit, Phys.\ Lett.\  B{\bf 658}, 170 (2008)
[arXiv:0706.4052 [hep-th]]; M.~Buric, D.~Latas, V.~Radovanovic and
J.~Trampetic, Phys.\ Rev.\  D{\bf 77} (2008) 045031 [arXiv:0711.0887 [hep-th]];
H.~Grosse and G.~Lechner, JHEP {\bf 0809} (2008) 131
[arXiv:0808.3459 [math-ph]].
\bibitem{neutral boson}
H. Grosse, and Y. Liao, Phys. Lett. B{\bf 520}, 63 (2001); H. Grosse, and Y. Liao, Phys. Rev. D{\bf 64}, 115007 (2001).
\bibitem{pho-neu}
P.~Schupp, J.~Trampetic, J.~Wess and G.~Raffelt, Eur.\ Phys.\ J.\  C{\bf 36}, 405 (2004)
[arXiv:hep-ph/0212292].
\bibitem{hez}
M.~Haghighat, M.~M.~Ettefaghi and M.~Zeinali, Phys. Rev. D{\bf 73}, 013007
(2006) [hep-ph/0511042].
\bibitem{eh}
M. Ettefaghi, and M. Haghighat, Phys. Rev. D{\bf 77}, 056009 (2008).
\bibitem{charge}
M.M. Sheikh-Jabbari, Phys. Rev. Lett. {\bf 84}, 5265 (2000).
\bibitem{phenomenology}
B.~Melic, K.~Passek-Kumericki and J.~Trampetic, Phys.\ Rev.\  D{\bf
72}, 054004 (2005)
[arXiv:hep-ph/0503133]; Phys.\ Rev.\  D{\bf 72} (2005) 057502
[arXiv:hep-ph/0507231]; A.~Alboteanu, T.~Ohl and R.~Ruckl, Phys.\ Rev.\
D{\bf 74} (2006) 096004 [arXiv:hep-ph/0608155]; A.~Alboteanu, T.~Ohl and
R.~Ruckl, Phys.\ Rev.\  D{\bf
76} (2007) 105018 [arXiv:0707.3595 [hep-ph]];
J.~Trampetic, Fortsch.\ Phys.\  {\bf 56}, 521 (2008)
[arXiv:0802.2030 [hep-ph]];
J.~A.~Conley and J.~L.~Hewett, [arXiv:0811.4218 [hep-ph]];
M.~Buric, D.~Latas, V.~Radovanovic and J.~Trampetic, Phys.\ Rev.\  D{\bf 75} (2007) 097701
[arXiv:hep-ph/0611299]; M. Haghighat, Phys. Rev. D{\bf 79}, 025011 (2009) [arXiv:0901.1069 [hep-ph]];
I.~Hinchliffe, N.~Kersting, and Y.~L.~Ma, Int. J. Mod. Phys. A{\bf
19}, 179 (2004);  M.~Haghighat and M.~M.~Ettefaghi, Phys. Rev. D{\bf
70}, 034017 (2004);  A.~Alboteanu, T.~Ohl and R.~Rückl, Phys. Rev. D{\bf 74}, 096004 (2006); M.~Mohammadi Najafabadi, Phys. Rev. D{\bf
74}, 025021 (2006);   M.~M.~Ettefaghi and M.~Haghighat, Phys. Rev. D{\bf 75}, 125002 (2007).
\bibitem{nucleosynthesis}
R.~Horvat and J.~Trampetic, arXiv:0901.4253 [hep-ph].
\bibitem{density}
S. Eidelman {\it et al.} [Particle Data Group Collaboration], Phys. Lett. B{\bf 592}, 1 (2004).
\bibitem{avar}
P. Gondolo and G. Gelmini, Nucl. Phys. B{\bf 360}, 145 (1991).
\end{thebibliography}
\end{document}